\documentclass[letter,twocolumn]{jpsj3}

\usepackage{txfonts}
\usepackage{url}
\usepackage{bm}
\usepackage{color}
\usepackage{enumitem}

\usepackage[whole]{bxcjkjatype}

\title{Fair sampling with temperature-targeted QAOA based on quantum--classical correspondence theory}

\author{Tetsuro Abe$^1$ and Shu Tanaka$^{1,2,3,4,5}$\thanks{shu.tanaka@keio.jp}}
\inst{
$^1$Graduate School of Science and Technology, Keio University, Kanagawa
223-8522, Japan \\
$^2$Department of Applied Physics and Physico-Informatics, Keio University, Kanagawa 223-8522, Japan \\
$^3$Keio University Sustainable Quantum Artificial Intelligence Center (KSQAIC), Keio University, Tokyo 108-8345,
Japan \\
$^4$Human Biology-Microbiome-Quantum Research Center (WPI-Bio2Q), Keio University, Tokyo 108-8345, Japan\\
$^5$Green Computing System Research Organization, Waseda University, Shinjuku, Tokyo 162-0042, Japan
} 

\abst{
In combinatorial optimization problems with degenerate ground states, fair sampling of degenerate solutions is essential. However, the quantum approximate optimization algorithm (QAOA) with a standard transverse-field mixer induces biases among degenerate states as circuit depth increases. Based on quantum--classical correspondence theory, we propose SBO-QAOA, which employs a temperature-dependent Hamiltonian encoding a Gibbs distribution as its ground state. Numerical simulations show that, unlike standard QAOA, SBO-QAOA yields ground-state probabilities converging to finite-temperature values with uniform distribution among degenerate states. These fairness and temperature-targeting properties are preserved even with only four variational parameters under a linear schedule.
}

\begin{document}
\maketitle

\textit{Introduction---}
Combinatorial optimization problems play an important role in a wide range of fields, including logistics, finance, drug discovery, and machine learning. 
In recent years, quantum technologies such as quantum annealing (QA) \cite{kadowaki1998quantum,tanaka2017quantum,tanahashi2019application} and the quantum approximate optimization algorithm (QAOA) \cite{farhi2014quantum, blekos2024review} have attracted attention as new metaheuristics for tackling these problems. 
In particular, for combinatorial optimization problems with degenerate ground states, finding a single optimal solution is sometimes insufficient. 
In such cases, the fair sampling of the set of degenerate ground states without bias becomes critically important.

In practical implementations of QA and QAOA, it is common to formulate the target problem as an Ising model and to employ a transverse-field term as a driver (mixer). 
However, both theoretical and experimental perspectives that transverse-field-driven QA and QAOA may fail to sample degenerate ground states fairly \cite{matsuda2009ground, tanaka2010roles, mandra2017exponentially, pelofske2025biased}. 
Indeed, previous studies have reported instances in which the distribution over degenerate ground states is biased toward specific solutions, even when the approximation ratio converges to unity.
This bias constitutes a fundamental limitation when these methods are used for sampling purposes. 
Various approaches such as specialized mixers \cite{bartschi2020grover, pelofske2021sampling, golden2022fair} and hybrid post-processing \cite{nakano2025fair} have been proposed, but they increase circuit complexity or require additional classical steps.
In addition, based on the principle of equal a priori probabilities in equilibrium statistical mechanics, it has also been shown that fair sampling can be achieved when simulated annealing is employed \cite{matsuda2009ground}.

In this study, we aim to achieve fair sampling in QAOA by focusing not on the mixer but on the design of the target Hamiltonian. 
Based on quantum–classical correspondence theory, Somma \textit{et al.} proposed a theoretical framework in which the thermal equilibrium state (Gibbs distribution) of a classical system is uniquely encoded as the ground state of a different quantum Hamiltonian \cite{somma2007quantum}. 
Yamamoto \textit{et al.} applied this framework to QA and demonstrated that fair sampling can be achieved through the effective Hamiltonian introduced by Somma \textit{et al.} \cite{yamamoto2020fair}. 
In the present work, we transfer this idea to QAOA, replacing the target Hamiltonian with an effective Hamiltonian while retaining the standard transverse-field mixer, thereby enabling both fair sampling and temperature-targeted sampling.

In this work, we refer to the effective Hamiltonian based on the theory proposed by Somma \textit{et al.} as the SBO Hamiltonian, and accordingly propose an SBO-QAOA in which this Hamiltonian is employed as the cost Hamiltonian in QAOA.
We apply both standard QAOA and SBO-QAOA to the same classical Ising model and numerically compare their sampling properties. 
In particular, we show that in SBO-QAOA the total probability of ground states converges to the value corresponding to the Gibbs distribution at a specified temperature. 
As a consequence, the probability distribution over the degenerate ground states becomes uniform. 
Furthermore, we demonstrate that this fair-sampling capability and temperature-targeting property are preserved even when parameter linearization is applied, reducing the number of variational parameters to just four.

\textit{Method and Experimental Setting---}
We consider an Ising model $\hat{H}_{0}$ for an $N$-spin system, defined as
\begin{equation}
    \hat{H}_{0}=-\sum_{1\le i<j \le N} J_{ij}\hat{\sigma}_i^{z}\hat{\sigma}_j^{z}-\sum_{i=1}^N h_i \hat{\sigma}_i^{z},
    \label{eq:H0}
\end{equation}
where $\hat{\sigma}_i^{z}$ denotes the Pauli $Z$ operator acting on spin $i$, $J_{ij}$ represents the interaction between spins $i$ and $j$, and $h_i$ is the longitudinal field applied to spin $i$. 
We define the classical energy for a computational basis state $|\sigma\rangle$ as $H_0(\sigma)=\langle\sigma|\hat{H}_0|\sigma\rangle$.

QAOA is a variational algorithm designed to find the ground state of the Hamiltonian $\hat{H}_{0}$.
In standard QAOA, the cost Hamiltonian is set to $\hat{H}_{\mathrm{C}}=\hat{H}_{0}$, and the mixer Hamiltonian is given by the transverse-field term $\hat{H}_{X}=\sum_{i=1}^{N}\hat{\sigma}_i^{x}$.
The initial state is taken to be $|+\rangle^{\otimes N}$.
The quantum state at circuit depth $p$ is given by
\begin{equation}
    |\psi_{p}\rangle=\prod_{k=1}^{p}e^{-i\beta_k \hat{H}_{X}}e^{-i\gamma_k \hat{H}_{\mathrm{C}}} |+\rangle^{\otimes N}.
    \label{eq:qaoa_state}
\end{equation}
The variational parameters are optimized by minimizing $\langle \psi_p|\hat H_C|\psi_p\rangle$ using the Powell method \cite{powell1964efficient}.

Next, we explain the theoretical construction of the target Hamiltonian used in SBO-QAOA. 
Within the framework of quantum–classical correspondence theory, reproducing the classical Gibbs distribution at temperature $T$, given by $P_{\mathrm{Gibbs}}(\sigma)\propto e^{-H_0(\sigma) /T}$, requires preparing a pure quantum state whose probability amplitudes are the square roots of the Boltzmann factors. 
\begin{equation}
    |\psi_\text{Gibbs}(T) \rangle = \frac{1}{\sqrt{Z(T)}} \sum_\sigma{e^{-{H_0(\sigma)}/{2T}} |\sigma \rangle},
\end{equation}
where $Z(T)=\sum_{\sigma}e^{-H_0(\sigma)/T}$ is the partition function of the Ising model at temperature $T$.
Throughout this paper, we set the Boltzmann constant to unity.

Somma \textit{et al.} constructed a method to define a quantum Hamiltonian that has $|\psi_\text{Gibbs}(T)\rangle$ as its unique ground state by assuming a classical Markov chain based on single-spin flips. The resulting Hamiltonian is given by
\begin{equation}
    \hat{H}_{S}(T)=-e^{-\alpha/T}\sum_{i=1}^{N}\left(\hat{\sigma}_i^{x}-e^{\hat{H}_{i} / T}\right),
    \label{eq:HS}
\end{equation}
where $\hat{H}_{i}$ denotes the local Hamiltonian consisting of the terms in $\hat{H}_{0}$ that depend on spin $i$, explicitly given by $\hat{H}_{i}=-\sum_{j(\neq i)}J_{ij}\hat{\sigma}_i^{z}\hat{\sigma}_j^{z}-h_i \hat{\sigma}_i^{z}$.
The coefficient $e^{-\alpha/T}$ on the right-hand side of Eq.~\eqref{eq:HS} is a scaling factor introduced to prevent the divergence of $\hat{H}_{S}(T)$ as $T\to 0$, where $\alpha = \max_i|\hat{H}_{i}|$. Here, $|\hat{H}_i|$ denotes the operator norm (maximum absolute eigenvalue).
In standard QAOA, the cost Hamiltonian is chosen as $\hat{H}_{\mathrm{C}}=\hat{H}_{0}$, whereas in SBO-QAOA we set $\hat{H}_{\mathrm{C}}=\hat{H}_{S}(T)$.

In this study, we consider two schemes for parameterizing the variational parameters $\{\gamma_k\}_{k=1}^{p}$ and $\{\beta_k\}_{k=1}^{p}$ in QAOA:
\begin{enumerate}[label=(\roman*)]
    \item Full-parameter: The parameters $\{\gamma_k\}_{k=1}^{p}$ and $\{\beta_k\}_{k=1}^{p}$ are treated as independent optimization variables, and a total of $2p$ parameters are optimized.
    \item Linearized: Following the parameter-reduction method proposed by Sakai \textit{et al.} \cite{sakai2024linearly}, each parameter is expressed as a linear function of the depth index:
        \begin{equation}
            \gamma_k=\gamma_\text{slope} \frac{k}{p}+\gamma_\text{intcp}, \quad
            \beta_k=\beta_\text{slope}\frac{k}{p}+\beta_\text{intcp}.
            \label{eq:linear_params}
        \end{equation}
        In this approach, the number of optimization variables is reduced to four, namely $(\gamma_\text{slope},\gamma_\text{intcp},\beta_\text{slope},\beta_\text{intcp})$, independent of the circuit depth $p$.
\end{enumerate}

The performance of QAOA depends strongly on the initial values of the variational parameters. 
Therefore, in this study we employ an initialization scheme based on Trotterized QA, obtained by discretizing a linear annealing schedule \cite{sack2021quantum}.
For depths $k=1,\dots,p$, the initial angles are given by
\begin{equation}
    \gamma_k^{(0)} = \frac{k}{p}\Delta t, \quad
    \beta_k^{(0)}  = \left(1-\frac{k}{p} \right)\Delta t,
    \label{eq:tqa_init_short}
\end{equation}
where $\Delta t$ is a scale parameter corresponding to the time step. In this work, we set $\Delta t=1$.

\begin{figure}[t]
    \centering
    \includegraphics[width=0.3\linewidth]{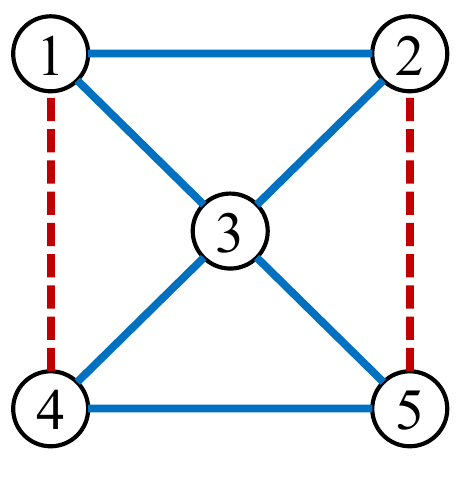}
    \caption{
        (Color online) The structure of the 5-spin toy model used in the numerical simulation.
        The nodes represent spins, and the edges represent interactions.
        Blue solid lines indicate ferromagnetic interactions ($J_{ij} =+1$), and red dashed lines indicate antiferromagnetic interactions ($J_{ij} = -1$).
        The competition between these interactions induces frustration, resulting in a six-fold ground-state degeneracy.
    }
    \label{fig:toy_model}
\end{figure}

For numerical validation, following Matsuda \textit{et al.} \cite{matsuda2009ground}, we employ a small toy model with degenerate ground states.
Specifically, we adopt an $N=5$ spin system, as shown in Fig.~\ref{fig:toy_model}. For all spins $i$, the longitudinal fields are set to $h_i=0$.
Owing to the frustration in the model shown in Fig.~\ref{fig:toy_model}, these six states share the same ground-state energy $E_0 = -4$.
We group these these six ground states into three pairs based on global spin-flip symmetry, referred to as state-pair 1, state-pair 2, and state-pair 3.
The explicit spin configurations are given below in the ket notation $|\sigma_1 \sigma_2 \sigma_3 \sigma_4 \sigma_5\rangle$:
\begin{align}
    \text{state-pair 1: }&|\uparrow\uparrow\uparrow\uparrow\uparrow\rangle,\ 
    |\downarrow\downarrow\downarrow\downarrow\downarrow\rangle,\label{eq:state1}\\
    \text{state-pair 2: }&|\uparrow\uparrow\uparrow\downarrow\downarrow\rangle,\ 
    |\downarrow\downarrow\downarrow\uparrow\uparrow\rangle,\label{eq:state2}\\
    \text{state-pair 3: }&|\uparrow\uparrow\downarrow\downarrow\downarrow\rangle,\ 
    |\downarrow\downarrow\uparrow\uparrow\uparrow\rangle\label{eq:state3}.
\end{align}
Here, $\uparrow$ and $\downarrow$ correspond to the eigenvalues $+1$ and $-1$, respectively. 
In this study, we evaluate sampling bias by tracking the probability distribution over these states.

Because the system size is small, the time-evolution operator is evaluated exactly as a full matrix without Trotter approximation.
From the resulting final state $|\psi_p\rangle$, we directly compute the probability $P_p(\sigma)=|\langle \sigma|\psi_p\rangle|^2$ in the computational basis $|\sigma\rangle$, which is then used to analyze the probability distribution over the degenerate ground states and to evaluate the distance from the Gibbs distribution.

\textit{Evaluation Metrics---}
In this study, we evaluate the performance of each method based on the probability distribution obtained by measuring the final state $|\psi_p\rangle$ in the computational basis $\{|\sigma\rangle\}$,
\begin{equation}
P_p(\sigma)=|\langle \sigma|\psi_p\rangle|^2.
\label{eq:Pp}
\end{equation}
We first define the set of degenerate ground states of the classical cost Hamiltonian $\hat{H}_0$ as $G = \{\sigma \mid H_0(\sigma) = E_0\}$.
The total probability of obtaining degenerate ground states is defined as
\begin{equation}
P_{\mathrm{GS}}=\sum_{\sigma\in G}P_p(\sigma).
\label{eq:Pgs_total}
\end{equation}

Because the instance considered in this study possesses global spin-flip symmetry, the degenerate ground states form pairs, as shown in Eqs.~\eqref{eq:state1}--\eqref{eq:state3}.
To visualize bias within the degenerate subspace, we define the probability of each state-pair as the sum of the probabilities of the two corresponding states:
\begin{align}
    P_{1}&=|\langle \uparrow\uparrow\uparrow\uparrow\uparrow|\psi_p\rangle|^2
          +|\langle \downarrow\downarrow\downarrow\downarrow\downarrow|\psi_p\rangle|^2,\\
    P_{2}&=|\langle \uparrow\uparrow\uparrow\downarrow\downarrow|\psi_p\rangle|^2
          +|\langle \downarrow\downarrow\downarrow\uparrow\uparrow|\psi_p\rangle|^2,\\
    P_{3}&=|\langle \uparrow\uparrow\downarrow\downarrow\downarrow|\psi_p\rangle|^2
          +|\langle \downarrow\downarrow\uparrow\uparrow\uparrow|\psi_p\rangle|^2.
\end{align}
We evaluate bias within the degenerate subspace using the state-pair probabilities $P_1$, $P_2$, and $P_3$, where the two states in each pair are symmetry equivalent.

Furthermore, to quantify the consistency with temperature-targeted sampling, we employ the total variation distance \cite{nelson2022high} between the Gibbs distribution $P_{\mathrm{Gibbs}}(\sigma)$ for $\hat{H}_0$ and the final distribution $P_p(\sigma)$:
\begin{equation}
D_{\mathrm{TVD}}=\frac{1}{2}\sum_{\sigma}\left|P_p(\sigma)-P_{\mathrm{Gibbs}}(\sigma)\right|.
\label{eq:Dtvd}
\end{equation}
A smaller value of $D_{\mathrm{TVD}}$ indicates that the final distribution is closer to the Gibbs distribution corresponding to the specified temperature $T$.

\begin{figure}[t]
    \centering
    \includegraphics[width=0.9\linewidth]{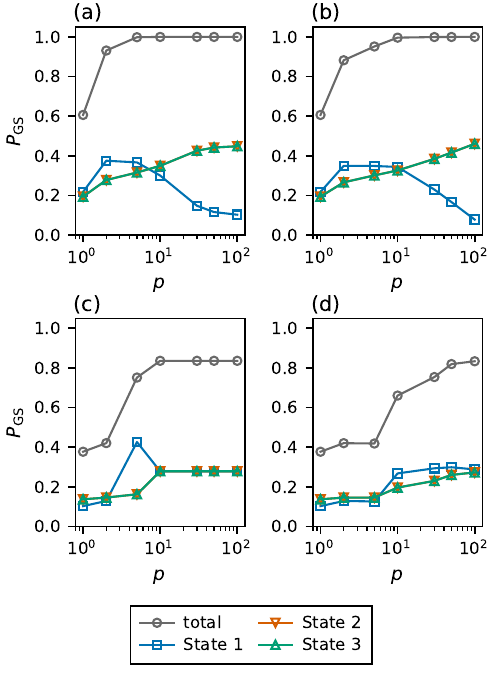}
    \caption{
    (Color online) Dependence of the ground-state probability on the circuit depth $p$. (a) full-parameter QAOA, (b) linearized QAOA, (c) full-parameter SBO-QAOA, and (d) linearized SBO-QAOA. The probability of each state-pair (state-pair 1–3) and the total ground-state probability $P_{\mathrm{GS}}$ are plotted.
    }
    \label{fig:probability_gs}
\end{figure}

\textit{Results---}
We first investigate the ground-state probability in the final distributions produced by QAOA and SBO-QAOA with $T=1$.
In Fig.~\ref{fig:probability_gs}, both the probability of each state-pair and the total ground-state probability $P_{\mathrm{GS}}$ (labeled ``total'') are shown.
The horizontal axis shows the circuit depth $p$ on a logarithmic scale, ranging from $p=1$ to $p=100$.

Figure~\ref{fig:probability_gs}(a) shows the results for full-parameter QAOA. As the circuit depth $p$ increases, the total ground-state probability rapidly increases, exceeding 0.9 within a few depths and approaching unity at $p \sim 10$. From the viewpoint of energy minimization, this indicates that increasing $p$ allows the algorithm to reach the degenerate ground-state manifold with high probability. However, the distribution within the degenerate subspace also changes: as $p$ increases, the probability of state-pair 1 initially increases but then decreases, whereas the probabilities of state-pairs 2 and 3 increase monotonically and eventually become dominant. Consequently, even in the high-$p$ regime where $P_{\mathrm{GS}}$ is close to unity, the probabilities $P_1$, $P_2$, and $P_3$ do not coincide, indicating a clear bias within the degenerate subspace.

Figure~\ref{fig:probability_gs}(b) presents the results for linearized QAOA. Linearized QAOA exhibits qualitatively the same biased behavior as full-parameter QAOA. This behavior demonstrates that even when the number of optimization variables is reduced from $2p$ in full-parameter QAOA to four, the probability distribution within the degenerate subspace does not automatically become uniform at large $p$ as long as the optimization targets the minimization of $\hat{H}_0$.

Figure~\ref{fig:probability_gs}(c) shows the results for full-parameter SBO-QAOA. Although $P_{\mathrm{GS}}$ increases with $p$, it does not reach unity, unlike in Figs.~\ref{fig:probability_gs}(a) and \ref{fig:probability_gs}(b). Instead, it saturates at approximately 0.83 in the large-$p$ regime. This saturation value is in close agreement with the ground-state probability of the Gibbs distribution at the specified temperature $T=1$, consistent with finite weight remaining on excited states. More importantly, the distribution within the degenerate subspace becomes increasingly uniform as $p$ increases: $P_1$, $P_2$, and $P_3$ approach one another and converge to nearly identical values at large $p$. Thus, when the SBO Hamiltonian is used as the target, the increase of $P_{\mathrm{GS}}$ is accompanied by a flattening of the probability distribution within the degenerate subspace, and fair sampling is numerically confirmed.

Figure~\ref{fig:probability_gs}(d) shows the results for linearized SBO-QAOA. The total ground-state probability $P_{\mathrm{GS}}$ increases monotonically with $p$ and converges to approximately 0.83 at large $p$, comparable to Fig.~\ref{fig:probability_gs}(c). The distribution within the degenerate subspace also becomes increasingly uniform as $p$ increases, and $P_1$, $P_2$, and $P_3$ eventually converge to similar values. The progression toward uniformity is smoother than in Fig.~\ref{fig:probability_gs}(c), and noticeable differences remain at small $p$. Nevertheless, fair sampling is numerically confirmed in the large-$p$ regime.

In summary, for both full-parameter QAOA and linearized QAOA targeting $\hat{H}_0$, increasing $p$ drives $P_{\mathrm{GS}}$ toward unity, while the probability distribution within the degenerate subspace does not become uniform and may even exhibit increasing bias. In contrast, for SBO-QAOA targeting the SBO Hamiltonian, $P_{\mathrm{GS}}$ saturates at a value corresponding to the Gibbs-distribution ground-state probability of $H_0$ at finite temperature, while the probabilities within the degenerate subspace become flat, approaching fair sampling.
Moreover, fair sampling is numerically confirmed even in linearized SBO-QAOA, where the optimization degrees of freedom are substantially reduced by restricting the number of variational parameters to four.

\begin{figure}[t]
    \centering
    \includegraphics[width=\linewidth]{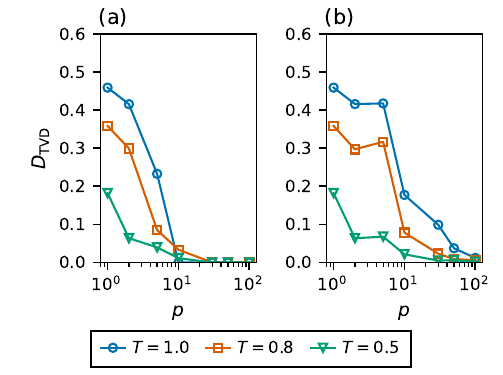}
    \caption{
    (Color online) Dependence of the total variation distance $D_{\mathrm{TVD}}$ between the final distribution $P_p(\sigma)$ and the target Gibbs distribution $P_{\mathrm{Gibbs}}(\sigma)$ on the circuit depth $p$. (a) Full-parameter SBO-QAOA and (b) linearized SBO-QAOA. Different curves correspond to different temperatures $T$.
    }
    \label{fig:TVD}
\end{figure}

Next, Fig.~\ref{fig:TVD} shows the dependence of the total variation distance $D_{\mathrm{TVD}}$ between the final distribution $P_p(\sigma)$ and the Gibbs distribution $P_{\mathrm{Gibbs}}(\sigma)$ on the circuit depth $p$ for several temperatures $T$. Figures~\ref{fig:TVD}(a) and \ref{fig:TVD}(b) present the results for full-parameter SBO-QAOA and linearized SBO-QAOA, respectively.

As shown in Fig.~\ref{fig:TVD}(a), $D_{\mathrm{TVD}}$ decreases with increasing $p$ for all temperatures and converges to values close to zero. 
This behavior indicates that SBO-QAOA does not merely concentrate probability on the degenerate ground states, but instead converges toward reproducing the entire state distribution corresponding to the specified temperature $T$.
A similar convergence behavior is observed for linearized SBO-QAOA, demonstrating that temperature-targeted sampling is preserved under parameter reduction as shown in Fig.~\ref{fig:TVD}(b).
%
%
Thus, fair sampling in SBO-QAOA emerges as a direct consequence of targeting the Gibbs distribution through the SBO Hamiltonian.

\textit{Conclusion---}
In this study, we compared standard QAOA and SBO-QAOA, which targets the SBO Hamiltonian $\hat{H}_\text{S}(T)$, under identical conditions for a five-spin toy model with degenerate ground states. We evaluated their properties from the viewpoints of fair sampling within the degenerate ground-state manifold and temperature-targeted sampling. We found that in full-parameter QAOA the ground-state probability $P_{\mathrm{GS}}$ approaches unity as the circuit depth $p$ increases, while the probability distribution within the degenerate ground states can become biased.

In contrast, for SBO-QAOA, $P_{\mathrm{GS}}$ converges to a value corresponding to finite temperature, and the probability distribution within the degenerate subspace tends to flatten. Furthermore, by evaluating the total variation distance $D_{\mathrm{TVD}}$ for SBO-QAOA, we showed that the final distribution systematically approaches the target Gibbs distribution as $p$ increases, demonstrating that fair sampling emerges as a direct consequence of the Gibbs-distribution-based design.

We also introduced a four-parameter linearization scheme, in which the depth-dependent angles are expressed as linear functions, and demonstrated that fair sampling in SBO-QAOA is preserved even when the number of variational parameters is reduced from $2p$ to four. These results suggest that fair sampling can be achieved without increasing the complexity of the mixer, by instead replacing the target Hamiltonian with $\hat{H}_{S}(T)$, and that this property can be maintained even with a small number of variational parameters.

An important issue to be addressed in future studies is scalability. In the present work, we evaluated the matrix exponential exactly for a small system. However, since $\hat{H}_{S}(T)$ generally contains terms of the form $\exp(\hat{H}_{i}/T)$, it can induce many-body interactions. Therefore, for direct implementation on gate-based quantum devices, it is necessary to develop efficient Pauli-string expansions and circuit decompositions of $\hat{H}_{S}(T)$, as well as low-order approximations tailored to the implementable interaction order. 

\section*{Acknowledgments}
This work was partially supported by the Japan Society for the Promotion of Science (JSPS) KAKENHI (Grant Number JP23H05447), the Council for Science, Technology, and Innovation (CSTI) through the Cross-ministerial Strategic Innovation Promotion Program (SIP), ``Promoting the application of advanced quantum technology platforms to social issues'' (Funding agency: QST), Japan Science and Technology Agency (JST) (Grant Number JPMJPF2221). The computations in this work were partially performed using the facilities of the Supercomputer Center, the Institute for Solid State Physics, The University of Tokyo. S. Tanaka wishes to express gratitude to the World Premier International Research Center Initiative (WPI), MEXT, Japan, for supporting the Human Biology-Microbiome-Quantum Research Center (Bio2Q).


\bibliographystyle{jpsj}
\bibliography{references}

@article{kadowaki1998quantum,
  title = {Quantum annealing in the transverse {I}sing model},
  author = {Kadowaki, Tadashi and Nishimori, Hidetoshi},
  journal = {Phys. Rev. E},
  volume = {58},
  issue = {5},
  pages = {5355--5363},
  numpages = {0},
  year = {1998},
  month = {Nov},
  publisher = {American Physical Society},
  doi = {10.1103/PhysRevE.58.5355},
  url = {https://link.aps.org/doi/10.1103/PhysRevE.58.5355}
}

@article{tanahashi2019application,
  title={Application of Ising machines and a software development for Ising machines},
  author={Tanahashi, Kotaro and Takayanagi, Shinichi and Motohashi, Tomomitsu and Tanaka, Shu},
  journal={J. Phys. Soc. Jpn.},
  volume={88},
  number={6},
  pages={061010},
  year={2019},
  publisher={The Physical Society of Japan}
}

@book{tanaka2017quantum,
  title={Quantum spin glasses, annealing and computation},
  author={Tanaka, Shu and Tamura, Ryo and Chakrabarti, Bikas K},
  year={2017},
  publisher={Cambridge University Press}
}

@incollection{tanaka2010roles,
  title={Roles of Quantum Fluctuation in Frustrated Systems--Order by Disorder and Reentrant Phase Transition},
  author={Tanaka, Shu and Hirano, M and Miyashita, S},
  booktitle={Quantum Quenching, Annealing and Computation},
  pages={215--234},
  year={2010},
  publisher={Springer}
}

@article{pelofske2025biased,
  title = {Biased degenerate ground-state sampling of small Ising models with converged quantum approximate optimization algorithm},
  author = {Pelofske, Elijah},
  journal = {Phys. Rev. E},
  volume = {111},
  issue = {5},
  pages = {054103},
  numpages = {10},
  year = {2025},
  month = {May},
  publisher = {American Physical Society},
  doi = {10.1103/PhysRevE.111.054103},
  url = {https://link.aps.org/doi/10.1103/PhysRevE.111.054103}
}

@article{yamamoto2020fair,
  title={Fair sampling by simulated annealing on quantum annealer},
  author={Yamamoto, Masayuki and Ohzeki, Masayuki and Tanaka, Kazuyuki},
  journal={J. Phys. Soc. Jpn.},
  volume={89},
  number={2},
  pages={025002},
  year={2020},
  publisher={The Physical Society of Japan}
}

@article{sakai2024linearly,
  title={Linearly simplified QAOA parameters and transferability},
  author={Sakai, Ryo and Matsuyama, Hiromichi and Tam, Wai-Hong and Yamashiro, Yu and Fujii, Keisuke},
  journal={arXiv:2405.00655},
}

@article{sack2021quantum,
  title={Quantum annealing initialization of the quantum approximate optimization algorithm},
  author={Sack, Stefan H and Serbyn, Maksym},
  journal={Quantum},
  volume={5},
  pages={491},
  year={2021},
  publisher={Verein zur F{\"o}rderung des Open Access Publizierens in den Quantenwissenschaften}
}

@article{nakano2025fair,
  title={Fair sampling of ground-state configurations using hybrid quantum-classical MCMC algorithms},
  author={Nakano, Yuichiro and Fujii, Keisuke},
  journal={arXiv:2512.14552},
}

@article{farhi2014quantum,
  title={A quantum approximate optimization algorithm},
  author={Farhi, Edward and Goldstone, Jeffrey and Gutmann, Sam},
  journal={arXiv:1411.4028},
}

@article{blekos2024review,
  title={A review on quantum approximate optimization algorithm and its variants},
  author={Blekos, Kostas and Brand, Dean and Ceschini, Andrea and Chou, Chiao-Hui and Li, Rui-Hao and Pandya, Komal and Summer, Alessandro},
  journal={Phys. Rep.},
  volume={1068},
  pages={1--66},
  year={2024},
  publisher={Elsevier}
}

@inproceedings{bartschi2020grover,
  title={Grover mixers for QAOA: Shifting complexity from mixer design to state preparation},
  author={B{\"a}rtschi, Andreas and Eidenbenz, Stephan},
  booktitle={Proc. IEEE Int. Conf. Quantum Computing and Engineering (QCE)},
  pages={72--82},
  year={2020},
  organization={IEEE}
}

@inproceedings{pelofske2021sampling,
  title={Sampling on nisq devices:" who’s the fairest one of all?"},
  author={Pelofske, Elijah and Golden, John and B{\"a}rtschi, Andreas and O'~Malley, Daniel and Eidenbenz, Stephan},
  booktitle={Proc. IEEE Int. Conf. Quantum Computing and Engineering (QCE)},
  pages={207--217},
  year={2021},
  organization={IEEE}
}

@article{golden2022fair,
  title={Fair sampling error analysis on NISQ devices},
  author={Golden, John and B{\"a}rtschi, Andreas and O’Malley, Daniel and Eidenbenz, Stephan},
  journal={ACM Trans. Quantum Comput.},
  volume={3},
  number={2},
  pages={1--23},
  year={2022},
  publisher={ACM New York, NY}
}

@article{somma2007quantum,
  title = {Quantum Approach to Classical Statistical Mechanics},
  author = {Somma, R. D. and Batista, C. D. and Ortiz, G.},
  journal = {Phys. Rev. Lett.},
  volume = {99},
  issue = {3},
  pages = {030603},
  numpages = {4},
  year = {2007},
  month = {Jul},
  publisher = {American Physical Society},
  doi = {10.1103/PhysRevLett.99.030603},
  url = {https://link.aps.org/doi/10.1103/PhysRevLett.99.030603}
}

@article{nelson2022high,
  title = {High-Quality Thermal Gibbs Sampling with Quantum Annealing Hardware},
  author = {Nelson, Jon and Vuffray, Marc and Lokhov, Andrey Y. and Albash, Tameem and Coffrin, Carleton},
  journal = {Phys. Rev. Appl.},
  volume = {17},
  issue = {4},
  pages = {044046},
  numpages = {20},
  year = {2022},
  month = {Apr},
  publisher = {American Physical Society},
  doi = {10.1103/PhysRevApplied.17.044046},
  url = {https://link.aps.org/doi/10.1103/PhysRevApplied.17.044046}
}

@article{matsuda2009ground,
  title={Ground-state statistics from annealing algorithms: quantum versus classical approaches},
  author={Matsuda, Yoshiki and Nishimori, Hidetoshi and Katzgraber, Helmut G},
  journal={New J. Phys.},
  volume={11},
  number={7},
  pages={073021},
  year={2009},
  publisher={IOP Publishing}
}

@article{mandra2017exponentially,
  title={Exponentially biased ground-state sampling of quantum annealing machines with transverse-field driving Hamiltonians},
  author={Mandra, Salvatore and Zhu, Zheng and Katzgraber, Helmut G},
  journal={Phys. Rev. Lett.},
  volume={118},
  number={7},
  pages={070502},
  year={2017},
  publisher={APS}
}

@article{powell1964efficient,
  title={An efficient method for finding the minimum of a function of several variables without calculating derivatives},
  author={Powell, Michael JD},
  journal={Comput. J.},
  volume={7},
  number={2},
  pages={155--162},
  year={1964},
  publisher={Oxford University Press}
}

\end{document}